\documentclass[prd,preprint,superscriptaddress,showpacs,byrevtex]{revtex4}
\usepackage{bm}
\def\fsl#1{\setbox0=\hbox{$#1$}           
   \dimen0=\wd0                                 
   \setbox1=\hbox{/} \dimen1=\wd1               
   \ifdim\dimen0>\dimen1                        
      \rlap{\hbox to \dimen0{\hfil/\hfil}}      
      #1                                        
   \else                                        
      \rlap{\hbox to \dimen1{\hfil$#1$\hfil}}   
      /                                         
   \fi}                                         %
\newcommand{\be}{\begin{equation}}
\newcommand{\ee}{\end{equation}}
\newcommand{\bea}{\begin{eqnarray}}
\newcommand{\eea}{\end{eqnarray}}
\newcommand{\beq}{\begin{equation}}
\newcommand{\eeq}{\end{equation}}
\newcommand{\beqs}{\begin{eqnarray}}
\newcommand{\eeqs}{\end{eqnarray}}

\newcommand{\aslash}{A\hspace{-0.067in}\slash}

\begin{document}
\title{ Lattice QCD Can Study Parton Distribution Function Inside Hadron}
\author{Gouranga C Nayak }\thanks{E-Mail: nayakg138@gmail.com}
%
%
\date{\today}
\begin{abstract}
Recently we have reported that the lattice QCD can not study the physical hadron formation from the unphysical quarks and gluons because it operates the unphysical QCD Hamiltonian of all the partons inside the hadron on the physical energy eigenstate of the hadron to obtain the physical energy eigenvalue of the hadron. However, since the parton distribution function (PDF) inside the hadron is unphysical (although it is well defined in QCD), we find in this paper that the unphysical energy of all the partons inside the hadron (instead of the physical energy of the hadron) can be used to study the PDF using the lattice QCD. Hence we find that the lattice QCD can study the parton distribution function inside the hadron from the first principle.
\end{abstract}
\maketitle
\pagestyle{plain}

\pagenumbering{arabic}

\section{Introduction}

Baryons (such as proton and neutron) and mesons (such as pion and kion) are collectively known as hadrons. A hadron is not a fundamental particle of the nature. The hadron is a composite particle consisting of quarks and gluons which are the fundamental particles of the nature.

The fundamental theory of the nature which describes the interaction between the quarks and gluons is known as the quantum chromodynamics (QCD). Similar to the quantum electrodynamics (QED) which is the quantum field theory of the classical electrodynamics, the QCD is the quantum field theory of the classical Yang-Mills theory \cite{ymb}. The classical electrodynamics is the U(1) gauge theory [or the abelian gauge theory] whereas the classical Yang-Mills theory is the SU(3) gauge theory [or the non-abelian gauge theory].

Unlike QED where the photons do not directly interact with each other the gluons interact directly with each other in QCD due to the non-abelian nature of the Yang-Mills theory. The gauge field lagrangian in QED contains quadratic power of the photon field whereas the gauge field lagrangian in QCD contains quadratic, cubic and quartic powers of the gluon field. Because of the presence of cubic and quartic powers of the gluon field in the QCD lagrangian the QCD is more complicated than QED and it becomes impossible to solve the full QCD analytically [see section \ref{path}].

After it was established that the QCD is renormalizable \cite{tvb}, the asymptotic freedom in QCD was found \cite{gwb,pob}. According to asymptotic freedom the renormalized QCD coupling becomes weak at the small distance and becomes strong at the large distance. Since the QCD coupling becomes weak at the small distance the short distance partonic level scattering cross section can be calculated by using the perturbative QCD (pQCD).

At the high energy colliders the short distance partonic scattering cross section is calculated by using the pQCD at fixed orders (such as at LO, NLO, NNLO etc.) in the coupling constant. Since we have not directly experimentally observed the quarks and gluons this short distance partonic scattering cross section is not a physical observable. Instead, what we directly experimentally measure at the high energy colliders is the hadronic cross section. Hence one finds that the quarks and gluons are unphysical but the hadrons are physical.

In order to calculate the physical hadronic cross section from the unphysical partonic scattering cross section at the high energy colliders one depends on the factorization theorem in QCD \cite{fcb,fcb1,fcb2}. According to the factorization theorem in QCD the partonic scattering cross section is folded with the parton distribution function (PDF) inside the hadron and with the parton to hadron fragmentation function (FF) to predict the hadron production cross section at the high energy colliders which is experimentally measured.

The parton distribution function (PDF) inside the hadron is a non-perturbative quantity in QCD. Hence the PDF can not be calculated by using the pQCD. For this reason the PDF is extracted from the experiments. Since the PDF is folded with the partonic cross section using the factorization theorem in QCD the PDF is well defined in QCD. The definition of the PDF in QCD is gauge invariant and is consistent with the factorization of the soft and collinear divergences in QCD at all orders in coupling constant \cite{csb,nkpb}.

As mentioned above since the PDF is a non-perturbative quantity in QCD it can not be calculated by using pQCD. Hence the PDF should be calculated from the first principle by using the non-perturbative QCD. However, the analytical solution of the non-perturbative QCD is not known. This is because of the presence of the cubic and quartic gluon fields in the QCD lagrangian which makes it impossible to perform the path integration in QCD analytically [see section \ref{path}].

For this reason the path integration in QCD is done numerically by using the lattice QCD method in the Euclidean time. In the study of the PDF in the lattice QCD method in the literature one uses the crucial equation \cite{ltb}
\bea
H^{\rm Partons}|E_n^{\rm Hadron}>=E_n^{\rm Hadron}|E_n^{\rm Hadron}>
\label{evq}
\eea
where $H^{\rm Partons}$ is the QCD Hamiltonian of all the partons inside the hadron, $|E_n^{\rm Hadron}>$ is the energy eigenstate of the hadron and $E_n^{\rm Hadron}$ is the energy eigenvalue of the hadron.

As mentioned above since the partons are not directly experimentally observed the QCD Hamiltonian $H^{\rm Partons}$ of all the partons inside the hadron is unphysical. Similarly since the hadron is directly experimentally observed the $|E_n^{\rm Hadron}>$ and $E_n^{\rm Hadron}$ are physical. Since the left hand side of eq. (\ref{evq}) is unphysical but the right hand side of eq. (\ref{evq}) is physical, one finds that the eq. (\ref{evq}) must be wrong, {\it i. e.},
\bea
H^{\rm Partons}|E_n^{\rm Hadron}>\neq E_n^{\rm Hadron}|E_n^{\rm Hadron}>.
\label{evqn}
\eea
Since the eq. (\ref{evq}) is used in the study of PDF in the lattice QCD method [see section \ref{pdf}] one finds from eq. (\ref{evqn}) that that the formulation of the PDF in the lattice QCD method is not consistent with the physics principle [see section \ref{pdfn}].

The eq. (\ref{evqn}) is consistent with the fact that the energy $E^{\rm Partons}(t)$ of all the partons inside the hadron given by
\bea
E^{\rm Partons}(t)=<H|H^{\rm Partons}|H>
\label{enh}
\eea
is not conserved and depends on time where $|H>=|E_0^{\rm Hadron}>$ is the energy eigenstate of the hadron $H$ in its ground state. The time dependence of $E^{\rm Partons}(t)$ in eq. (\ref{enh}) is due to the non-vanishing boundary surface term [non-zero energy flux $E_{\rm flux}(t)$] in QCD which arises due to the confinement of quarks and gluon inside the finite size hadron \cite{nkb}. In fact because of the confinement of the quarks and gluons inside the finite size hadron, the total conserved energy in QCD is given by $E^{\rm Partons}(t)+E_{\rm flux}(t)$ where \cite{nkb}
\bea
\frac{d}{dt} [E^{\rm Partons}(t)+E_{\rm flux}(t)]=0.
\label{csv}
\eea

From eq. (\ref{enh}) we find
\bea
H^{\rm Partons}|H>=E^{\rm Partons}(t)|H>
\label{enh0}
\eea
which is consistent with eq. (\ref{evqn}) because the left hand side and the right hand side of eq. (\ref{enh0}) are unphysical. Note that unlike eq. (\ref{evq}) [where the left hand side of eq. (\ref{evq}) is unphysical but the right hand side of eq. (\ref{evq}) is physical], since the left hand side and the right hand side of eq. (\ref{enh0}) are unphysical the eq. (\ref{enh0}) is consistent in QCD whereas the eq. (\ref{evq}) is not consistent in QCD.

Therefore the correct equation in QCD due to the confinement of quarks and gluons inside the finite size hadron is the eq. (\ref{enh0}) which should be used instead of eq. (\ref{evq}) to study the parton distribution function inside the hadron using the lattice QCD method.

Because of eq. (\ref{evqn}) we have recently reported in \cite{nklmb} that the lattice QCD can not study the physical hadron formation from the unphysical quarks and gluons because it operates the unphysical QCD Hamiltonian of all the partons inside the hadron on the physical energy eigenstate of the hadron to obtain the physical energy eigenvalue of the hadron [see also section \ref{pdfn}].

In this paper we find that, since the parton distribution function (PDF) inside the hadron is unphysical (although it is well defined in QCD), the unphysical energy $E^{\rm Partons}(t)$ of all the partons inside the hadron as given by eq. (\ref{enh}) [instead of the physical energy $E_0^{\rm Hadron}$ of the hadron as given by eq. (\ref{evq})] can be used to study the PDF using the lattice QCD. Hence we find that the lattice QCD can study the parton distribution function inside the hadron from the first principle.

The paper is organized as follows. In section II we discuss the lattice QCD method which is used in the literature to study the parton distribution function inside the hadron. In section III we show the inconsistency in this lattice QCD method to study the PDF. In section IV we present the consistent formulation of the lattice QCD method to study the parton distribution function inside the hadron. Section V contains conclusions.

\section{ Lattice QCD Method To Study The Parton Distribution Function Inside The Hadron }

The parton distribution function (PDF) of the quark inside the hadron $H$ is given by \cite{csb,nkpb}
\bea
f_q(\xi)=\frac{1}{4\pi} \int_{-\infty}^{+\infty} dy^- e^{-i\xi P^+y^-} <H|{\bar \psi}(y^-) \gamma^+ [{\cal P} e^{-ig T^c\int_0^{y^-} dy'^-A^{+c}(y'^-)}] \psi(0)|H>.
\label{qdb}
\eea
where $\psi(x)$ is the quark field, $A_\mu^a(x)$ is the gluon field, $P^+$ is the momentum of the hadron, the $|H>$ is the energy eigenstate of the hadron $H$ and $\xi$ is the momentum fraction of the quark inside the hadron with respect to the momentum of the hadron.

Since the analytical solution of the non-perturbative QCD is not known the lattice QCD employs the numerical method to calculate the PDF inside the hadron.
To study the PDF inside the hadron $H$ the lattice QCD computes the vacuum expectation value of the non-perturbative two point correlation function $<\Omega|{\cal O}_H^{\rm Partons}(x){\cal O}_H^{\rm Partons}(0)|\Omega>$ of the partonic operator ${\cal O}_H^{\rm Partons}(x)$ and the vacuum expectation value of the non-perturbative three point correlation function $<\Omega|{\cal O}_H^{\rm Partons}(x){\cal P}(x'){\cal O}_H^{\rm Partons}(0)|\Omega>$ of the partonic operators ${\cal O}_H^{\rm Partons}(x)$ and ${\cal P}(x)$. The $|\Omega>$ is the non-perturbative QCD vacuum state which is different from the pQCD vacuum state $|0>$. The partonic operators ${\cal O}_H^{\rm Partons}(x)$ and ${\cal P}(x)$ are defined as follows.

\subsection{ Partonic Operator For The Hadron Formation }

In order to form the hadron $H$ from the partons the partonic operator ${\cal O}_H^{\rm Partons}(x)$ is chosen in such a way that it carries the same quantum numbers of the hadron. For example, for the proton formation from the quarks the partonic operator ${\cal O}_H^{\rm Partons}(x)$ in $<\Omega|{\cal O}_H^{\rm Partons}(x){\cal O}_H^{\rm Partons}(0)|\Omega>$ and in $<\Omega|{\cal O}_H^{\rm Partons}(x){\cal P}(x'){\cal O}_H^{\rm Partons}(0)|\Omega>$ is given by
\bea
{\cal O}_P^{\rm Partons}(x)=\epsilon_{ijk} \psi^u_i(x) \psi^u_j(x) C\gamma_5 \psi^d_k(x)
\label{pop}
\eea
where $C$ is the charge conjugation operator and $u,~d$ represent up, down quarks. Similarly for the neutron formation from the quarks the partonic operator ${\cal O}_H^{\rm Partons}(x)$ in $<\Omega|{\cal O}_H^{\rm Partons}(x){\cal O}_H^{\rm Partons}(0)|\Omega>$ and in $<\Omega|{\cal O}_H^{\rm Partons}(x){\cal P}(x'){\cal O}_H^{\rm Partons}(0)|\Omega>$ is given by
\bea
{\cal O}_N^{\rm Partons}(x)=\epsilon_{ijk} \psi^d_i(x) \psi^d_j(x) C\gamma_5 \psi^u_k(x).
\label{pon}
\eea

\subsection{ Parton Distribution Function Operator }

The parton distribution function operator in the quantum field theory should be consistent with the number operator definition of the parton. In addition to this the parton distribution function operator also contains the gauge link [see eq. (\ref{qdb})] which makes the PDF gauge invariant and consistent with the factorization of soft and collinear divergences at all orders in coupling constant \cite{csb,cssb,nkpb,nkpb1}. From eq. (\ref{qdb}) we find that the quark parton distribution function operator
is given by
\bea
{\cal P}(x,t) = {\psi}^\dagger(0,t) [{\cal P}e^{-igT^c \int_0^x dx'_\mu A^{\mu c}(x')}] \psi(x,t).
\label{pdob}
\eea

\subsection{ Path Integration In QCD and The Vacuum Expectation Value of The Non-Perturbative Correlation Function Of Partonic Operators }\label{path}

The vacuum expectation value of the two-point non-perturbative correlation function $<\Omega|{\cal O}_H^{\rm Partons}(x){\cal O}_H^{\rm Partons}(0)|\Omega>$ of the partonic operator ${\cal O}_H^{\rm Partons}(x)$ in QCD is given by \cite{mtb,abb}
\bea
&& <\Omega|{\cal O}_H^{\rm Partons}(x'){\cal O}_H^{\rm Partons}(x'')|\Omega>= \int [dA] [d{\bar \psi}] [d\psi]~ {\cal O}_H^{\rm Partons}(x'){\cal O}_H^{\rm Partons}(x'')~{\rm det}[\frac{\delta C_f^a}{\delta \omega^b}]~\nonumber \\
&& \times {\rm exp}[i\int d^4x [-\frac{1}{4}F_{\mu \sigma}^b(x) F^{\mu \sigma b}(x)-\frac{1}{2\alpha}[C_f^a(x)]^2+{\bar \psi}(x)[(i{\not \partial} -m)+gT^b\aslash^b(x)]\psi(x)]]
\label{veb}
\eea
and the vacuum expectation value of the three-point non-perturbative correlation function $<\Omega|{\cal O}_H^{\rm Partons}(x){\cal P}(x'){\cal O}_H^{\rm Partons}(0)|\Omega>$ of the partonic operators ${\cal O}_H^{\rm Partons}(x)$ and ${\cal P}(x)$ in QCD is given by \cite{mtb,abb}
\bea
&& <\Omega|{\cal O}_H^{\rm Partons}(x'){\cal P}(x''){\cal O}_H^{\rm Partons}(x''')|\Omega>= \int [dA] [d{\bar \psi}] [d\psi] ~{\cal O}_H^{\rm Partons}(x'){\cal P}(x''){\cal O}_H^{\rm Partons}(x''')~{\rm det}[\frac{\delta C_f^a}{\delta \omega^b}]~\nonumber \\
&& \times {\rm exp}[i\int d^4x [-\frac{1}{4}F_{\mu \sigma}^b(x) F^{\mu \sigma b}(x)-\frac{1}{2\alpha}[C_f^a(x)]^2+{\bar \psi}(x)[(i{\not \partial} -m)+gT^b\aslash^b(x)]\psi(x)]]
\label{veb3}
\eea
where $C_f^a(x)$ is the gauge fixing term, $\alpha$ is the gauge fixing parameter and
\bea
F_{\mu \sigma}^b(x) = \partial_\mu A_\sigma^b(x) - \partial_\sigma A_\mu^b(x) + gf^{bdh} A_\mu^d(x) A_\sigma^h(x).
\label{fmb}
\eea
In eqs. (\ref{veb}) and (\ref{veb3}) we do not have any ghost fields because we directly work with the ghost determinant ${\rm det}[\frac{\delta C_f^a}{\delta \omega^b}]$. The typical choice for the gauge fixing term in pQCD at the high energy colliders is the covariant gauge fixing $C_f^a(x) =\partial^\sigma A_\sigma^a(x)$. Note that there is a normalization factor $\frac{1}{Z[0]}$ in eqs. (\ref{veb}) and (\ref{veb3}) but since they cancel in the ratio in eq. (\ref{phc}) we have not included it here where $Z[0]$ is the generating functional in QCD in the absence of external sources.

From eq. (\ref{fmb}) one finds that the exponentials in eqs. (\ref{veb}) and (\ref{veb3}) contain cubic and quartic powers of the gluon field $A_\mu^b(x)$ which makes it impossible to perform the path integrations in eqs. (\ref{veb}) and (\ref{veb3}) analytically. This is the main reason why the non-perturbative QCD can not be studied analytically to study the PDF inside the hadron.

Since the path integrations in eqs. (\ref{veb}) and (\ref{veb3}) can not be done analytically, the lattice QCD performs the path integrations in eqs. (\ref{veb}) and (\ref{veb3}) numerically in the Euclidean time.

\subsection{ Lattice QCD Method To Study The Parton Distribution Function Inside The Hadron }\label{pdf}

The ratio of the three-point to two-point correlation functions to study the PDF in the lattice QCD is given by
\bea
R(t,t') = \frac{\sum_{\vec x} \sum_{{\vec x}'} e^{i{\vec P}\cdot ({\vec x}-{\vec x}')}~<\Omega|{\cal O}_H^{\rm Partons}(t,{\vec x}){\cal P}(t',{\vec x}'){\cal O}_H^{\rm Partons}(0)|\Omega>}{\sum_{\vec x} e^{i{\vec P}\cdot {\vec x}}<\Omega|{\cal O}_H^{\rm Partons}(t,{\vec x}){\cal O}_H^{\rm Partons}(0)|\Omega>}
\label{phc}
\eea
where ${\vec P}$ is the momentum of the hadron. Inserting complete set of energy eigenstates of the hadron
\bea
\sum_n |E_n^{\rm Hadron}><E_n^{\rm Hadron}|=1
\label{phg}
\eea
in eq. (\ref{phc}) we find
\bea
&&R(t,t') = [ \frac{1}{\sum_n <\Omega|{\cal O}_H^{\rm Partons}(0) e^{itH^{\rm Partons}}|E_n^{\rm Hadron}><E_n^{\rm Hadron}|{\cal O}_H^{\rm Partons}(0)|\Omega>}] \nonumber \\
&&\times ~[\sum_n \sum_{n'} <\Omega|{\cal O}_H^{\rm Partons}(0) e^{itH^{\rm Partons}}|E_n^{\rm Hadron}><E_n^{\rm Hadron}|e^{-it'H^{\rm Partons}} {\cal P}(0) e^{it'H^{\rm Partons}}|E_{n'}^{\rm Hadron}>\nonumber \\
&&\times ~<E_{n'}^{\rm Hadron}|{\cal O}_H^{\rm Partons}(0)|\Omega>]
\label{phh}
\eea
where $H^{\rm Partons}$ is the QCD Hamiltonian of all the partons inside the hadron $H$.

Using eq. (\ref{evq}) in (\ref{phh}) we find
\bea
&&R(t,t') = [\frac{1}{ \sum_n <\Omega|{\cal O}_H^{\rm Partons}(0) |E_n^{\rm Hadron}><E_n^{\rm Hadron}|{\cal O}_H^{\rm Partons}(0)|\Omega>e^{-tE_n^{\rm Hadron}}}] \nonumber \\
&&\times ~[\sum_n \sum_{n'} <\Omega|{\cal O}_H^{\rm Partons}(0)|E_n^{\rm Hadron}> <E_n^{\rm Hadron}| {\cal P}(0)|E_{n'}^{\rm Hadron}><E_{n'}^{\rm Hadron}|{\cal O}_H^{\rm Partons}(0)|\Omega> \nonumber \\
&& \times ~e^{-tE_n^{\rm Hadron}}e^{-t'[E_{n'}^{\rm Hadron}-E_n^{\rm Hadron}]}].
\label{phjf}
\eea
where $E_n^{\rm Hadron}$ is the energy of the hadron $H$ in its n$th$ level.

Neglecting all the higher level energy contributions and keeping only the ground state of the hadron we find from eq. (\ref{phjf}) [in the limit $t>>>t'$ as $t,t' \rightarrow \infty$]
\bea
&&[R(t,t')]_{t>>>t' \rightarrow \infty} = \frac{1}{<\Omega|{\cal O}_H^{\rm Partons}(0) |H><H|{\cal O}_H^{\rm Partons}(0)|\Omega>e^{-tE^{\rm Hadron}}} \nonumber \\
&&<\Omega|{\cal O}_H^{\rm Partons}(0)|H> <H| {\cal P}(0)|H><H|{\cal O}_H^{\rm Partons}(0)|\Omega> e^{-tE^{\rm Hadron}}e^{-t'[E^{\rm Hadron}-E^{\rm Hadron}]} \nonumber \\
\label{phkf}
\eea
where $|H>$ is the ground state energy eigenstate $|H>=|E_0^{\rm Hadron}>$ of the hadron $H$ and $E$ is the ground state energy eigenvalue $E=E_0^{\rm Hadron}$ of the hadron $H$.

From eq. (\ref{phkf}) we find
\bea
&&[R(t,t')]_{t>>>t' \rightarrow \infty}  = <H| {\cal P}(0)|H>.
\label{phnf}
\eea
Hence from eqs. (\ref{phc}) and (\ref{phnf}) we find that the moment of the parton distribution function (PDF) $<H| {\cal P}(0)|H>$ inside the hadron $H$ is given by
\bea
<H| {\cal P}(0)|H>=[\frac{\sum_{\vec x} \sum_{{\vec x}'} <\Omega|{\cal O}_H^{\rm Partons}(t,{\vec x}){\cal P}(t',{\vec x}'){\cal O}_H^{\rm Partons}(0)|\Omega>}{\sum_{\vec x} <\Omega|{\cal O}_H^{\rm Partons}(t,{\vec x}){\cal O}_H^{\rm Partons}(0)|\Omega>}]_{t>>>t' \rightarrow \infty}
\label{phof}
\eea
where the right hand side of this equation is computed by using the lattice QCD.

\section{ Inconsistency In The Lattice QCD Method To Study The Parton Distribution Function Inside The Hadron }\label{pdfn}

However, the correct equation in QCD due to the confinement of quarks and gluons inside the finite size hadron \cite{nkb} is given by eq. (\ref{evqn}) [not eq. (\ref{evq})]. Hence using eq. (\ref{evqn}) in (\ref{phh}) we find
\bea
&&R(t,t') \neq [\frac{1}{ \sum_n <\Omega|{\cal O}_H^{\rm Partons}(0) |E_n^{\rm Hadron}><E_n^{\rm Hadron}|{\cal O}_H^{\rm Partons}(0)|\Omega>e^{-tE_n^{\rm Hadron}}}] \nonumber \\
&&\times ~[\sum_n \sum_{n'} <\Omega|{\cal O}_H^{\rm Partons}(0)|E_n^{\rm Hadron}> <E_n^{\rm Hadron}| {\cal P}(0)|E_{n'}^{\rm Hadron}><E_{n'}^{\rm Hadron}|{\cal O}_H^{\rm Partons}(0)|\Omega> \nonumber \\
&& \times ~e^{-tE_n^{\rm Hadron}}e^{-t'[E_{n'}^{\rm Hadron}-E_n^{\rm Hadron}]}].
\label{phjfn}
\eea
which does not agree with eq. (\ref{phjf}).

\section{ Consistent Formulation of The Lattice QCD Method To Study The Parton Distribution Function Inside The Hadron }

We found from eq. (\ref{phjfn}) that the eq. (\ref{evq}) can not be used in (\ref{phh}) to study the parton distribution function inside the hadron. As mentioned in the introduction the correct equation in QCD due to the confinement of quarks and gluons inside the finite size hadron is eq. (\ref{enh0}) [not eq. (\ref{evq})]. Hence the eq. (\ref{enh0}) should be used in eq. (\ref{phh}) [instead of eq. (\ref{evq})] to study the parton distribution function inside the hadron in the lattice QCD method.

Eq. (\ref{enh0}) for the n$th$ level becomes
\bea
H^{\rm Partons}|E_n^{\rm Hadron}>=E_{n}^{\rm Partons}(t)|E_n^{\rm Hadron}>.
\label{phi}
\eea
Using eq. (\ref{phi}) in (\ref{phh}) we find in the Euclidean time
\bea
&&R(t,t') =[ \frac{1}{ \sum_n <\Omega|{\cal O}_H^{\rm Partons}(0) |E_n^{\rm Hadron}><E_n^{\rm Hadron}|{\cal O}_H^{\rm Partons}(0)|\Omega>e^{-tE_{n}^{\rm Partons}(t)}}] \nonumber \\
&&\times ~[\sum_n \sum_{n'} <\Omega|{\cal O}_H^{\rm Partons}(0)|E_n^{\rm Hadron}> <E_n^{\rm Hadron}| {\cal P}(0)|E_{n'}^{\rm Hadron}><E_{n'}^{\rm Hadron}|{\cal O}_H^{\rm Partons}(0)|\Omega>\nonumber \\
&&e^{-tE_{n}^{\rm Partons}(t)}e^{-t'[E_{n'}^{\rm Partons}(t')-E_{n}^{\rm Partons}(t')]}].
\label{phj}
\eea
where $E_{n}^{\rm Partons}(t)$ is the total energy of all the partons inside the hadron $H$ in its n$th$ level.

Neglecting all the higher level energy contributions and keeping only the ground state of the hadron we find from eq. (\ref{phj}) [in the limit $t>>>t'$ as $t,t' \rightarrow \infty$]
\bea
&&[R(t,t')]_{t>>>t' \rightarrow \infty} = \frac{1}{<\Omega|{\cal O}_H^{\rm Partons}(0) |H><H|{\cal O}_H^{\rm Partons}(0)|\Omega>e^{-tE^{\rm Partons}(t)}} \nonumber \\
&&<\Omega|{\cal O}_H^{\rm Partons}(0)|H> <H| {\cal P}(0)|H><H|{\cal O}_H^{\rm Partons}(0)|\Omega>\nonumber \\
&&e^{-tE^{\rm Partons}(t)}e^{-t'[E^{\rm Partons}(t')-E^{\rm Partons}(t')]}
\label{phk}
\eea
where $|H>=|E_0^{\rm Hadron}>$ of the hadron $H$ in its ground state and $E^{\rm Partons}(t)=E_0^{\rm Partons}(t)$ is the total energy of all the partons inside the hadron $H$ in its ground state.

From eq. (\ref{phk}) we find
\bea
&&[R(t,t')]_{t>>>t' \rightarrow \infty}  = <H| {\cal P}(0)|H>.
\label{phn}
\eea
Hence from eqs. (\ref{phc}) and (\ref{phn}) we find that the moment of the parton distribution function (PDF) $<H| {\cal P}(0)|H>$ inside the hadron $H$ is given by eq. (\ref{phof}) where the right hand side of eq. (\ref{phof}) is computed by the lattice QCD.

From eqs. (\ref{phof}), (\ref{phn}) and (\ref{phi}) we find that since the parton distribution function (PDF) inside the hadron is unphysical (although it is well defined in QCD), the unphysical energy $E^{\rm Partons}(t)$ of all the partons inside the hadron as given by eq. (\ref{enh}) [instead of the physical energy $E_0^{\rm Hadron}$ of the hadron as given by eq. (\ref{evq})] can be used to study the PDF using the lattice QCD.

Hence we find that the lattice QCD can study the parton distribution function inside the hadron from the first principle.

\section{Conclusions}
Recently we have reported that the lattice QCD can not study the physical hadron formation from the unphysical quarks and gluons because it operates the unphysical QCD Hamiltonian of all the partons inside the hadron on the physical energy eigenstate of the hadron to obtain the physical energy eigenvalue of the hadron. However, since the parton distribution function (PDF) inside the hadron is unphysical (although it is well defined in QCD), we have found in this paper that the unphysical energy of all the partons inside the hadron (instead of the physical energy of the hadron) can be used to study the PDF using the lattice QCD. Hence we have found that the lattice QCD can study the parton distribution function inside the hadron from the first principle.

\end{document}